\documentclass[a4paper]{jpconf}

\usepackage{graphicx}
\usepackage{xspace}
\usepackage{amsmath}
\usepackage{xcolor}
\usepackage{mathtools}
\usepackage[normalem]{ulem}

\usepackage{aasjournals}

\usepackage{url}

\newcommand{\msun}{$M_\odot$\xspace}

\newcommand{\zsun}{$Z_\odot$\xspace}

\newcommand{\numberspace}{\ensuremath{\;}}
\newcommand{\nsp}{\numberspace}

\newcommand{\Msun}{\ensuremath{M_\odot}}
\newcommand{\Lsun}{$L_\odot$\xspace}

\newcommand{\etar}{\eta^{\phantom 2}_{R}}
\newcommand{\etab}{\eta^{\phantom 2}_{B}}

\begin{document}

\title{Quantification of Incertitude in Black Box Simulation Codes}

\author{A~C~Calder$^{1,2}$, M~M~Hoffman$^3$,
D~E~Willcox$^1$, M~P~Katz$^{4}$, F~D~Swesty$^1$, and S Ferson$^5$}

\address{$^1$ Department of Physics and Astronomy, 
Stony Brook University, Stony Brook, NY 11794-3800, USA}
\address{$^2$ Institute for Advanced Computational Science,
Stony Brook University, Stony Brook, NY 11794-5250, USA}
\address{$^3$ National Radio Astronomy Observatory, Charlottesville VA, 22902, USA}
\address{$^4$ NVIDIA Corporation, 2788 San Tomas Expressway, Santa Clara, CA, 95050 USA}
\address{$^5$ Institute for Risk and Uncertainty, University of
Liverpool L69 7ZF, UK}

\ead{alan.calder@stonybrook.edu}

\begin{abstract}
We present early results from a study addressing the question of how one
treats the propagation of incertitude, that is, epistemic uncertainty, 
in input parameters in astrophysical simulations.  As an example, we 
look at the propagation of incertitude in control parameters for stellar 
winds in MESA stellar evolution simulations.
We apply two methods
of incertitude propagation, the Cauchy Deviates method and the
Quadratic Response Surface method, to quantify the output uncertainty 
in the final white
dwarf mass given a range of values for wind parameters. The methodology 
we apply is applicable to the problem of propagating input incertitudes
through any simulation code treated as a ``black box," 
i.e.\ a code for which the algorithmic details are either inaccessible
or prohibitively complicated. We have made the tools developed for 
this study freely available to the community.
\end{abstract}

\section{Introduction}
Much of theoretical astrophysics relies on large-scale simulation for
progress because phenomena of interest typically incorporate multiple
physical processes interacting over a wide range of scales. Even study
of the basic processes, e.g.\ plasma and fluid dynamics, requires
simulations that can challenge extant computer architectures.
Because of both the importance of simulation and the dynamic nature of
state-of-the-art computer architectures, considerable resources go
into the development and maintenance of simulation codes. Many of
these advanced codes, built with person-years of development
effort, are made publicly available and are used currently by many
investigators in addition to the original developers.

Despite the capabilities of modern codes and architectures, uncertainty
imposes limits on the results of simulations. This issue is often 
neglected in studies in part because of the difficulty in addressing it.
As with verification and validation, uncertainty quantification is
critical to credible computational science, particularly in astrophysics
where most simulation results are predictions, that is, a description
of the state of a system under conditions for which the computational
model has not been validated~\cite{roache1998verification,
oberkampf2010verification,1998aiaa}. 

There are two types of uncertainty to consider, aleatory
uncertainty (variability)  and epistemic uncertainty, also
known as  incertitude. Aleatory uncertainties occur when
there is a variability or inherent randomness, and examples
in stellar astrophysics include initial mass and composition
among stars. Aleatory uncertainty is described by
a (possibly unknown) distribution.  In contrast, incertitude is a 
lack of knowledge about some aspect of the model and is a
consequence of incomplete information,
including a paucity or absence of data or an incomplete theoretical
understanding.

The research presented here addresses the question of how to propagate
incertitude in the input parameters though a complex simulation
instrument while treating it as a ``black box." Doing so avoids the
possibly difficult process of analytically propagating uncertainty through the
individual arithmetic operations involved in the calculation by rules of interval
arithmetic, which may require more expertise than a typical code user
has. The simulation instrument we employ in this study is the MESA
stellar evolution code, and we treat it as a black box, applying it to
the evolution of 1 \msun ZAMS stars with minimal tuning.  This
paper presents initial results using relatively low resolution
simulation data aimed at showcasing the method. The complete study is
forthcoming~\cite{hoffmanetal2017}.

\section{Incertitude propagation techniques}

The goal of incertitude propagation is to quantify uncertainty in the
output of a simulation given ranges in input parameters representing
their incertitudes. There are a variety of techniques for formally
propagating incertitude via interval arithmetic, but most of these
involve instrumenting or modifying a particular code or algorithm,
which is not possible when addressing a black box code. 
In this study, we apply methods of incertitude 
propagation appropriate for black box codes, two simple
sampling schemes and two more advanced techniques.

\subsection{Sampling the Space}

The simplest approach to exploring the bounds of a black box
given incertitude in the input parameters is to sample the
parameter space. Two obvious approaches are to uniformly 
sample it and to randomly sample it from a uniform 
distribution. We applied both of these sampling methods,
and include the results from the random sample from a uniform 
distribution below.

If the black box is known to be a linear function, then
sampling each uncertain parameter at its upper and lower 
bound will provide the bounds on the output. Such an approach
is limited, however, by what is known as the ``curse of 
dimensionality," which we discuss below.
If the black box is not known to be a linear function, then both
a uniform sample and a sample randomly drawn from a uniform
distribution serve to characterize the output and can guide
investigators in selecting more advanced methods of uncertainty
quantification to apply. In principle, a large random sample
would allow establishing bounds of the output, but in practice
such an approach is limited by the possibility of pathological
points in the underlying function and the curse of dimensionality.

\subsection{The Cauchy Deviates Method}
The Cauchy distribution (also referred to as the Lorentz 
distribution or the Breit-Wigner distribution) is described by 
the probability distribution
\begin{equation}
  \label{eq:Cauchy}
p(z) = \frac{\Delta}{\pi(z^2+\Delta^2)} \; ,
\end{equation}
where $\Delta$ is the scale parameter of the distribution and 
$z$ is the random variable. The method of Cauchy Deviates 
(CD) relies on properties of 
this distribution, particularly that a linear combination of independent 
Cauchy-distributed variables is itself Cauchy distributed. 
The method proceeds by associating the width of the uncertainty 
interval for each input
variable with the scale parameter $\Delta$ in equation~\ref{eq:Cauchy}. 
Then, by sampling Cauchy-distributed inputs, one is able to
obtain an interval for the output of the model, which,
by the linear combination property, is also Cauchy distributed.
One then extracts the scale parameter from the output distribution, 
$\Delta_y$, which is associated with the width
of the uncertainty just as with the input parameters. Details 
of this method may be found in
\cite{kreinovich:2004} and complete details of our implementation may
be found in \cite{hoffmanetal2017}.

It is important to note that in
employing the Cauchy distribution {\em we make no assumption that the
input parameters are Cauchy distributed}.  This assertion may seem
counter-intuitive.  The method simply uses the fact that in a linear 
black-box, the input Cauchy distributions will be stretched and translated 
into an output Cauchy distribution.  By comparing how the full-width half
maxima (FWHM) of the input distributions, set by the scale parameters
$\Delta_i$ for each of the $n$ input distributions, are distorted into
the FWHM for the output distribution, characterized by $\Delta_y$, one 
can quantify how the 
black box translates and stretches input incertitude intervals into 
output intervals.   The Cauchy Deviates method cannot predict the 
distribution of the model outputs as one does not know the distribution
of the inputs.  Thus the use of the input Cauchy distribution is merely
a tool to quantitatively measure how much the black box function 
alters input intervals into output intervals.

The CD technique works for a linear functionality, providing
a way to understand uncertainty propagation through a black box for an
approximately
linear function. As our study typified, astrophysical simulations are
commonly nonlinear, and our approach to handling this nonlinear problem
is to tile the input intervals into
regions that are suspected to be nearly linear over that area and then apply
the CD technique on each region. 

It is possible to measure the uncertainty of this $\Delta_y$ and create
confidence intervals, i.e., one can calculate the statistical
uncertainty in our incertitude interval from the number of 
samples of the black box, $N$. One can relate the number of samples
to the interval through
\begin{equation}
  \sigma_{\Delta_y} = \Delta_y \cdot \sqrt{\frac{2}{N}}.
\end{equation}
Using this relationship one can calculate the number of samples
necessary to construct a confidence interval with a specific,
statistical uncertainty, $\sigma$. One can then say
at a chosen level of confidence that the true interval
lies within our constructed Cauchy Deviates interval, plus or minus $\sigma$.

Because of this relationship between the confidence and the number
of samples $N$, the CD method can constrain the uncertainty
for a black box regardless of how large $n$ is. This property of
the method becomes more powerful as the number of uncertain inputs grows. 

\subsection{The Quadratic Response Surface Method}

The Quadratic Response Surface Method (QRSM) addresses the problem
of a nonlinear function by constructing a quadratic response
surface that is fit to the output of the black box function that has
been sampled over the domain of the input parameter space.
The method thus avoids the need for tiling as is the case with the
CD method. Also, despite being nonlinear, an astute choice
of interval and fitting function allows determination
of the final interval with a reasonable number of samples. 

The QRSM method proceeds by first fitting a
quadratic surface to the output of the black box function that has
been sampled over the domain of the input parameter space.  Once a
quadratic surface has thus been obtained, the maximum and minimum of
the surface can be found in order to obtain an estimate of the bounds
on the output of the black box given the chosen ranges of input uncertainty.
Finding the maximum and minimum of the quadratic response surface over
the incertitude domain is essentially a constrained optimization problem,
but finding the maximum and minimum of an
$n$-dimensional quadratic function over the rectangular $n$-dimensional
domain is challenging. We take the approach of  \cite{kreinovich:2008}
and use an ellipsoidal domain method, which allows for the use of a 
relatively simple mathematical algorithm for finding optima on an 
ellipsoidal domain.

\begin{figure}[h]
	\centering
  \includegraphics[angle=0,
    width=4.0in]{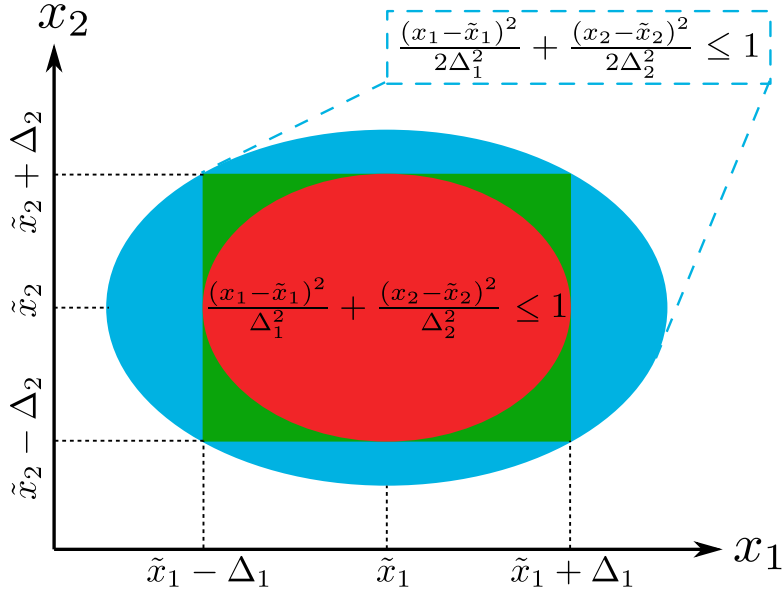}
  \caption{Ellipses constructed that either inscribe and circumscribe
    the original rectangular interval. We can construct
	ellipses from our square interval using the pictured equations. 
	For a number of uncertain inputs $n>2$, the ellipse is generalized to
    a multivariate ellipsoid.}
  \label{fig:ells}
\end{figure}

As the input parameter space is rectangular, an obvious question that
arises in moving to an elliptical domain is whether or not the
elliptical domain should inscribe or circumscribe the rectangular domain. 
Figure~\ref{fig:ells} illustrates the two possibilities.
The choice for the ellipse to inscribe or circumscribe the rectangle depends principally
on whether or not it is safe to exclude the corner regions of the
rectangular parameter domain.
If one is confident that the 
jointly extreme combinations of the input variables do not occur
in the modeled system, as might be tenable if one assumes their 
uncertainties are independent, then the inscribed ellipse is safe, 
as it is a subset of the rectangular uncertainty domain.
If one is not sure that the maxima and minima will not occur on the
corners of the input uncertainty domain, the safe path is to select
the circumscribed ellipse and ensure that the optimization domain is at least
as large as that within the rectangle.
One caveat, here, follows from areas within the circumscribed
ellipsoid being outside of the $n$-dimensional rectangle. These
regions of parameter space may not be physically realizable, which
could potentially lead to an unrealistic optimization solution,
for which the black box function may be difficult to evaluate.

The ellipsoid serves as a constraint on the model, and there
are many possible circumscribing ellipsoids for a given rectangle.
We chose to minimize the volume of the 
$n$-dimensional circumscribing ellipsoids for a given
$n$-dimensional rectangle and performed the fit accordingly. 
Ultimately, the goal is to find the maximum and minimum of an
objective function, which for the QRSM method is a quadratic function given by
\begin{equation} \label{eq:obfun}
  f(x_1,...,x_n) = f_0 + \sum_{i=1}^n f_i \cdot x_i \ 
  + \sum_{i=1}^n \sum_{j=1}^n f_{i,j}\cdot x_i \cdot x_j
\end{equation}
that approximates the output of the black box model over the domain of
the input parameters.  Equation~(\ref{eq:obfun}) describes a
multivariate quadratic model, and one can obtain the coefficients
$f_0$, $f_i$, and $f_{i,j}$ by obtaining a fit to the output data
using multivariate quadratic regression analysis.  The output of the
black-box model is then approximately bounded by the maximum and
minimum of equation~(\ref{eq:obfun}) over the ellipsoidal regions in the
input parameter domain.

\section{Case Study: the MESA Simulation Instrument}
The instrument we treat as our black box code for this study
is MESA (Modules for Experiments in Stellar Astrophysics)
\cite{MESA:I,MESA:II,MESA:III}. The MESA code is
open source and is widely used in astrophysical research. It
is an excellent example of a community code that is the culmination
of years of development. As might be expected from 
incorporating multiple physical effects on multiple scales, 
the algorithms in MESA are complicated and defy the 
relatively easy analysis needed to 
analytically propagate incertitude in input parameters.


We applied the two advanced incertitude propagation techniques to the 
problem of evolving stars with MESA. Specifically, we performed suites of simulations
evolving zero age main sequence (ZAMS) stars through their lives
until they become C/O white dwarfs. 
The simulations were based on a setup available in the MESA test
suite, \texttt{1M\_pre\_ms\_to\_wd},
with models of 1 \Msun\ with a solar
metallicity, \zsun = 0.02, and the stars were evolved to a luminosity
of 0.1\Lsun, the endpoint of stellar evolution.

Stellar winds are the mechanism of mass loss during the evolution of
stars, but a first-principles understanding of these remains elusive.
Instead, descriptions are parameterized by stellar properties such as
luminosity, mass, and radius.  MESA implements a number of
prescriptions for mass loss during evolution on the red giant branch
(RGB) and asymptotic giant branch (AGB) \cite{reimers1975,1975psae.book..229R,bloecker1995,
dejager1988}.  For simplicity in this study, we chose models 
of the winds during these phases that do not assume a temperature dependence.

The two epistemically uncertain parameters for these winds
are the Reimers and
Bl{\"o}ckers scaling factors, $\etar$ and $\etab$, corresponding
to RGB winds and AGB winds respectively.  These parameters describe 
the RGB mass loss rate~\cite{1975psae.book..229R} and AGB mass 
loss rate~\cite{bloecker1995}, as
\begin{equation} \label{eq:Rwind}
  \dot{M}_{R} = 4 \cdot 10^{-13}\text{  } \etar\text{  } 
  \frac{LR}{M}\text{    }[M_{\odot}\text{  yr}^{-1}]
  \end{equation}
and
\begin{equation} \label{eq:Bwind}
  \dot{M}_{B} = 1.932 \cdot 10^{-21}\text{  } \etab M^{-3.1}\cdot 
  L^{3.7}R \text{    }[M_{\odot}\text{  yr}^{-1}] \; .
\end{equation}
Here $L$, $M$, and $R$ are the luminosity, mass, and radius of the 
star in solar units.
The scaling factors, $\etar$ and
$\etab$, describe the strength of the winds and 
MESA expresses these through the
controls \texttt{Reimers\_scaling\_factor} and
\texttt{Blocker\_scaling\_factor}, respectively.

\section{Results}

In this section, we describe performing suites of simulations with
MESA and applying the methods of uncertainty quantification. The results 
for each suite are a range of white dwarf masses presenting the bounds on
the mass given the incertitude of our input.

\begin{figure}[h]
        \centering
  \includegraphics[angle=0,
    width=4.5in]{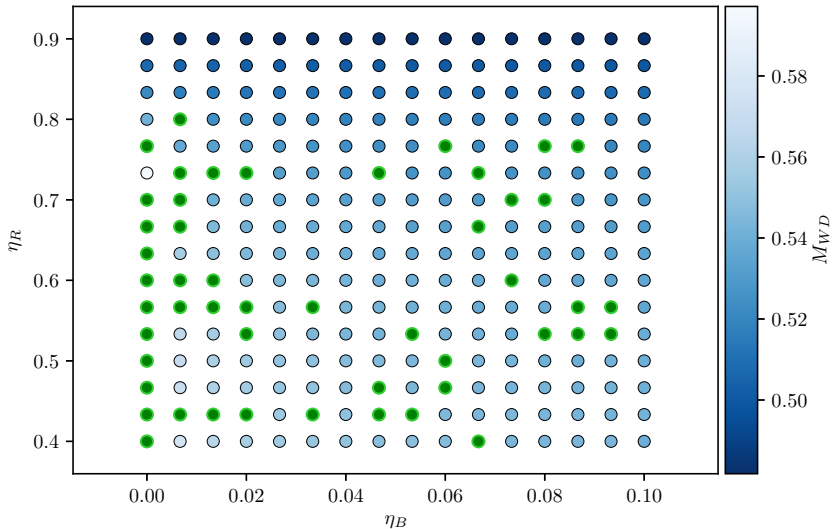}
  \caption{
    Initial regular
    grid of 256 $1.0M_{\odot}$ ZAMS models. The axes show 
    $\protect\etar$ vs.\ $\protect\etab$,
    and the  colormap indicated the
    final CO WD star masses.  Models for which the
	simulations did not complete are highlighted in
	 green.} 
  \label{fig:initial1}
\end{figure}

We began our study with a regularly sampled uniform grid with 
$\etab$ sampled 
between 0.0 and 0.1, and  $\etar$ was sampled
between 0.4 and 0.9. These values were chosen following
an earlier study  of 
$1.0M_{\odot}$ ZAMS mass stars with similar evolution tracks in
\cite{karakas:2016} and \cite{pignatari:2016}.
The results are shown in 
figure~\ref{fig:initial1}.
Points at $\etab=0.0$ failed to complete for lower
values of $\etar$, as seen in the figure.
After this initial study, the domain was slightly adjusted so
that $\etab$ was sampled regularly
between 0.01 and 0.1  and
$\etar$ was sampled regularly between
0.3 and 0.9.  In addition to an initial exploration,
this suite of regularly sampled
simulations also served as a baseline with which to compare the CD and
and QRSM methods.
We note that any robust method for incertitude propagation must be able to 
accommodate the realistic situation that simulation codes are unable
to carry out simulations for all values of input parameters for a variety
of reasons. In this regard, MESA is a typical proxy for many other 
astrophysical simulation codes.

\begin{figure}[h]
        \centering
  \includegraphics[angle=0,
    width=4.5in]{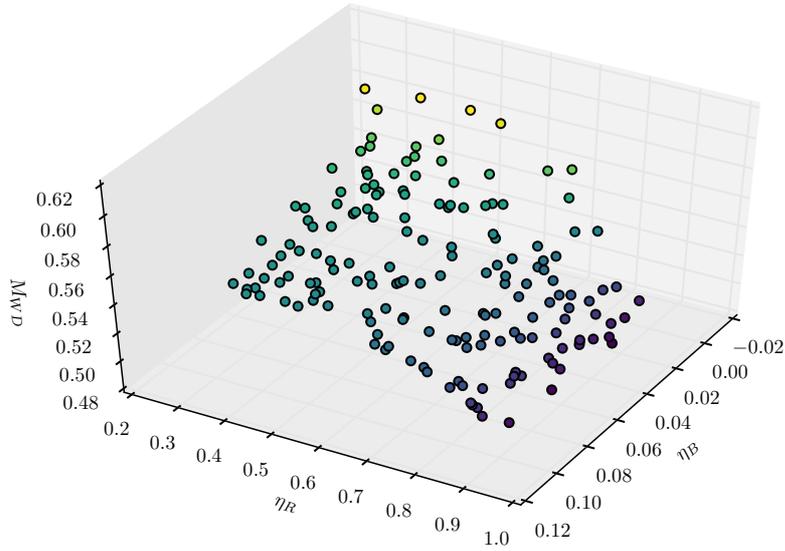}
  \caption{
    A suite of 200 models with values of
    $\protect\etar$ and $\protect\etab$ chosen at random from a uniform
    distribution between $\protect\etar=[0.3,0.9]$ and
	$\protect\etab=[0.01,0.1]$.}
  \label{fig:initial2}
\end{figure}

We next performed a suite of 200 simulations randomly sampled from a
uniform distribution. In this case, results from simulations that did
not complete are omitted. This suite better illustrated the parameter
space, particularly the obvious nonlinearity. Figure~\ref{fig:initial2} 
depicts the results of this suite of simulations.
For this suite, the minimum sampled mass is 0.4902 \Msun,
and the maximum sampled mass is 0.5984 \Msun.

\begin{figure}[h]
        \centering
  \includegraphics[angle=0,
    width=4.5in]{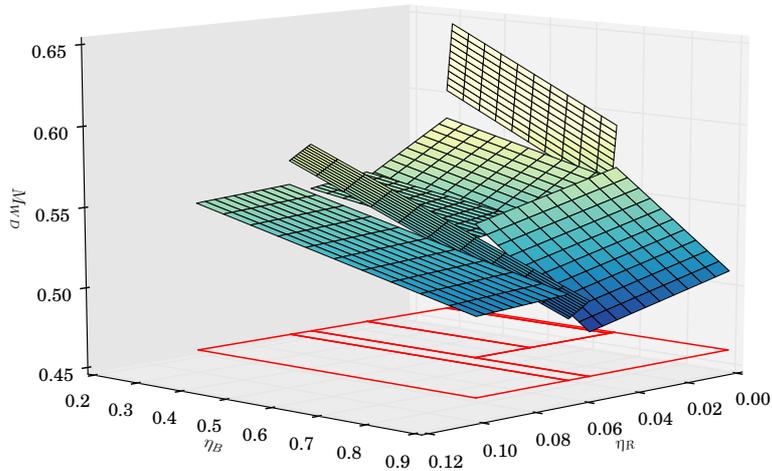}
  \caption{
	Parameter space broken up into 6 linear planes for
	application of the CD approach to uncertainty quantification. Linear tiles are determined
        via a region-growing algorithm that fits planes to samples of 
	the parameter space subject to a constraint of the coefficient
	 of determination for each plane fit.}
  \label{fig:nonlinear1}
\end{figure}

The obvious nonlinearity observed in the first two suites of
simulations indicated modification of the CD method would be required. As
explained by \cite{kreinovich:2004}, a way to deal with non-linearity
is to subdivide the parameter space into linear sections and for
each calculate a set of Cauchy-distributed inputs 
within the intervals. When bounds are found for each
section, the maximum of the maxima and the minimum of the
minima produce the bounds for the full space. 
Figure~\ref{fig:nonlinear1} illustrates the resultant tiles. 
For the suites of
simulations that went into the CD study, we found an interval 0.5395
$\pm$ 0.0542 \Msun\ with a standard deviation of 0.0077 \Msun, giving
the range as [0.4776, 0.6014] [0.4699, 0.6091] [0.4622, 0.6168]
\Msun\ with 68.3\%, 99.5\%, and 99.7\% confidence, respectively.  

The objective function found for our QRSM study is given by
\begin{align}\label{eq:quadfit}
  \begin{aligned}
  M_{\rm WD} = f(\etab,\etar) =\nsp & \nsp0.5596 - 1.1323~\etab + 0.1378~\etar\\
  & + 5.7688~\eta_B^{2} - 0.2060~\eta_R^{2} \\
  & + 0.4646~\etab\cdot \etar \; ,
  \end{aligned}
\end{align}
where the coefficients $f_0$,
$f_i$, and $f_{i,j}$ in equation \ref{eq:obfun} have been replaced by
the actual values obtained by fitting (see figure \ref{fig:nonlinear2} below).  
For the purpose of obtaining the
objective function, the sampling was done over the actual incertitude
domain instead of on either the inner or outer ellipse.  The
computation of the output incertitude range based on the objective
function is, however, found using the elliptical constraints.
Figure~\ref{fig:nonlinear2} illustrates the quadratic regression fit.
 \begin{figure}
        \centering
  \includegraphics[angle=0,
    width=4.5in]{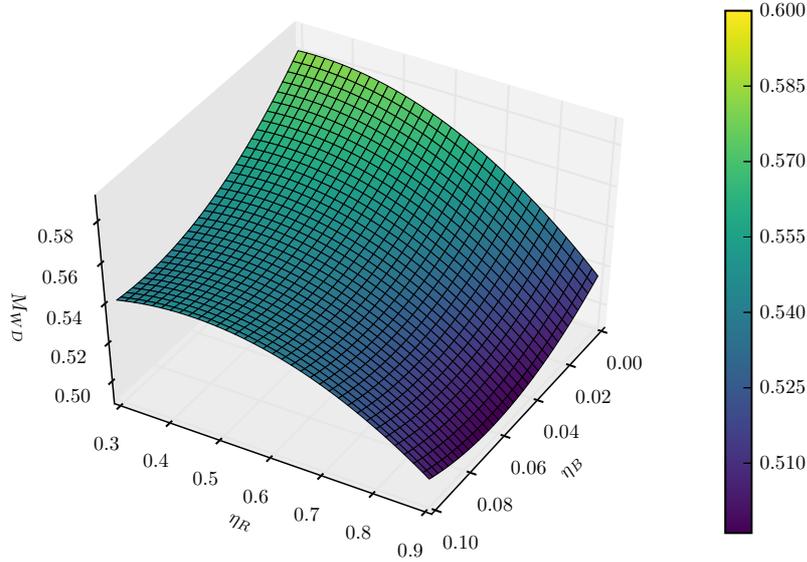}
  \caption{Surface plot of the multivariate quadratic regression
        fit. Note that the maximum and minimum
    are on the edges---there are no local extrema within the space.}
  \label{fig:nonlinear2}
\end{figure}
For our QRSM study, we found ranges [0.4979,0.5606] \Msun\ and
[0.4811,0.6119] \Msun\ for the inscribed and circumscribed ellipses,
respectively.

\section{Discussion and Conclusions}

The purpose of this manuscript is to present the methodology
for a study of incertitude propagation in black box simulation
codes, and the results presented here, ranges of white dwarf masses, 
are not definitive. We presented this study as a look at work
in progress and defer final calculated ranges to the complete
work \cite{hoffmanetal2017}.
As may be seen in the results from the first uniform grid of
points in the input parameter domain, many of the MESA simulations
did not converge. While it is possible to produce more robust
simulations with MESA by ``tuning" the input to improve convergence,
we specifically did not do this so as to keep our black box 
completely consistent for all simulations. The complete study
will review the issue of convergence of the simulations.
Our conclusion is that these methods, CD and QRSM, allow the
formal propagation of the input incertitude into incertitude intervals
of the simulation outputs in a straight forward
and computationally tractable way. 

In this case study, we utilized only
two variables with incertitude. The full advantage of the advanced methods 
that we demonstrated, CD and QRSM, does not become apparent until
applied to higher dimensional incertitude problems. 
One might be tempted to conclude that since the case study problem can be
broken up into linear sub regions that one could easily evaluate the
model at the extrema of the incertitude ranges. While this could be easily 
done for the two-dimensional case, in a situation with $n$ uncertain input
parameters, this approach would require $2^n$ model evaluations to estimate
the output uncertainty of the model. This is the so-called curse of dimensionality.
In contrast, the advanced methods require far fewer model evaluations in 
order to estimate the model output uncertainty.

Another comparison we wish to make is between the CD and QRSM methods. The
CD method has an advantage in that one has some knowledge of how many model
sample points are needed in order to specify the desired confidence in
the estimate of the model output uncertainty. In contrast, with the QRSM method
there is no statistical knowledge of how many sample points are needed in
order assess the precision with which the model output incertitude is known.
The QRSM method, however, does posses two advantages over the CD method. First,
the QRSM method makes no assumption of linearity and does not require subdividing
the input parameter incertitude intervals into quasi-linear  sub-intervals,
thereby avoiding computationally costly model evaluations needed to evaluate
the degree of linearity. Second, the QRSM method potentially can require far
fewer model evaluations in that as few as  $n(n-1)/2+2n+1$ are needed to determine
the coefficients of the ellipsoid given in equation \ref{eq:obfun}.
In a situation where $n$ is large, this property could result in a significant 
advantage realized through the use of the  QRSM method.
One can only gain confidence in the estimates of the model output uncertainty,
however,
by carrying out a convergence test where the number of sample points is
increased. The minimum number of points  yields an exact output uncertainty only in the case of
a perfectly quadratic model.

Finally, we wish to point out a limitation on the interpretation of the output model 
response. This response cannot be physically interpreted as a probability distribution because the 
distribution of input parameters is unknown by definition of incertitude. All that
can be determined in the case of input incertitudes is a model output uncertainty.
Were the distributions of the input uncertainties known, the output response could
have an interpretation as a probability distribution. The methods described
here, however, are not suitable to address this problem. Research into uncertainty quantification
for the problem of known distributions of input uncertainties, however, is underway.


The source code for the tools applied to this study, Star Simulation 
Techniques for Research in Uncertainty Quantification, is freely 
available at \url{https://github.com/StarSTRUQ}. The site
includes the software for generating the Cauchy deviates samples,
performing the QRSM analysis, and the tiling, which subdivides a sampled
domain into planar tiles constrained by the planar fit.

\ack

This work was supported in part by the US Department of Energy under grant
DE-FG02-87ER40317 and by the US National Science Foundation under grant
AST-1211563. The research described here originated with co-author Melissa
Hoffman's thesis at Stony Brook University. The simulations performed
for this study were performed with the MESA stellar evolution code and
the authors thank its developers and community for making this resource
available. The authors would like to thank Stony Brook Research Computing
and Cyberinfrastructure, and the Institute for Advanced Computational
Science at Stony Brook University for access to the high-performance LIred
and SeaWulf computing systems, the latter of which was made possible by a
\$1.4M National Science Foundation grant (\#1531492). The authors also thank
Tracy M.\ Calder for previewing the manuscript. This research has made
use of NASA's Astrophysics Data System Bibliographic Services.  

\section*{References}


\end{document}